# Applying economic measures to lapse risk management with machine learning approaches

Stéphane Loisel[*1], Pierrick Piette[†1,3,4] and Cheng-Hsien Jason Tsai[‡2]

[1]Univ Lyon, Université Claude Bernard Lyon 1, Institut de Science Financière et d'Assurances (ISFA), Laboratoire SAF EA2429, F-69366, Lyon, France

[2]National Chengchi University (NCCU), College of Commerce, Department of Risk Management and Insurance, Taipei, Taiwan

[3]Sorbonne Université, CNRS UMR 8001, Laboratoire de Probabilités Statistique et Modélisation, LPSM, 4 place Jussieu, F-75005 Paris, France.

[4]Sinalys, 6 rue de Téhéran, 75008 Paris, France

**Abstract**

Modeling policyholders lapse behaviors is important to a life insurer since lapses affect pricing, reserving, profitability, liquidity, risk management, as well as the solvency of the insurer. Lapse risk is indeed the most significant life underwriting risk according to European Insurance and Occupational Pensions Authority's Quantitative Impact Study QIS5. In this paper, we introduce two advanced machine learning algorithms for lapse modeling. Then we evaluate the performance of different algorithms by means of classical statistical accuracy and profitability measure. Moreover, we adopt an innovative point of view on the lapse prediction problem that comes from churn management. We transform the classification problem into a regression question and then perform optimization, which is new for lapse risk management. We apply different algorithms to a large real-world insurance dataset. Our results show that XGBoost and SVM outperform CART and logistic regression, especially in terms of the economic validation metric. The optimization after transformation brings out significant and consistent increases in economic gains.

**Keywords:** lapse; machine learning; SVM; XGBoost; life insurance

**JEL Classification:** C52; C53; G22.

[*] Email: stephane.loisel@univ-lyon1.fr. The author acknowledges support from research chair DAMI (Data Analytics and Models for Insurance) sponsored by BNP Paribas Cardif.
[†] Email: pierrick.piette@gmail.com.
[‡] Email: ctsai@nccu.edu.tw. The author is grateful to the Ministry of Science and Technology of Taiwan (project number MOST 105-2410-H-004 -070 -MY3) and Risk and Insurance Research Center of National Chengchi University for the financial supports.

# 1. Introduction

Lapse risk is the most significant risk associated with life insurance when compared with longevity risk, expenses risk, and catastrophe risk. Policyholders of life insurance may choose to surrender their policies at any time for cash values, or opt to stop paying premiums and leave policies to become invalid eventually. Lapses have significant impacts on the profitability, or even on the solvency, of a life insurer as many studies demonstrate. They may reduce expected profits (Hwang and Tsai, 2018), cause underwriting expenses unrecovered (Tsai et al., 2009; Pinquet et al., 2011), impair the effectiveness of an insurer's asset-liability management (Kim, 2005a; Eling and Kochanski, 2013) and bring in liquidity threats as experienced by US life insurers in the late 1980s.

When lapses vary with interest rates as suggested by Dar and Dodds (1989), Kuo et al. (2003), Kim (2005b, 2005c) and Cox and Lin (2006), they become even more detrimental to life insurers (Tsai et al., 2009). Many papers argue that the option to surrender a policy for the cash value might account for a large proportion of the policy value, e.g., Albizzati and Geman (1994), Grosen and Løchte Jørgensen (2000), Bacinello (2003), Bauer et al. (2006), Gatzert and Schmeiser (2008), and Consiglio and Giovanni (2010). The above reasoning and finding may be the reasons why the fifth Quantitative Impact Study (QIS5), conducted by the European Insurance and Occupational Pensions Authority (EIOPA) in 2011 regarding the implementation of Solvency II, reports that lapse risk accounts for about 50% of the life underwriting risks.

The significance of lapse risk draws attentions of scholars to study what causes policyholders to lapse their policies. We may classify the literature into being macro- or micro-oriented. Macro-oriented papers (e.g., Dar and Dodds, 1989; Kuo et al., 2003; Kim, 2005b, 2005c; Cox and Lin, 2006) focus on how lapse rates (the proportion of lapsed policies to the total number of sampled policies within a period of time) are affected by environmental variables such as interest rates, unemployment rates, gross domestic product, and returns in capital markets, as well as by company characteristics like size and organizational form.

Micro-oriented papers secure data from insurers on individual policies to investigate the determinants of the lapse propensities/tendencies. The identified determinants include the characteristics of policyholders and the features of life insurance products/policies (see Renshaw and Haberman (1986), Kagraoka (2005), Cerchiara et al. (2005), Milhaud et al. (2011), Pinquet et al. (2011), and Eling and Kochanski (2013) among others.). Eling and Kochanski (2013) and Campbell et al. (2014) provide extensive reviews of the literature on lapses.[1]

This paper extends the micro-oriented line of literature in two ways. Firstly, we introduce machine learning algorithms including Extreme Gradient Boosting (XGBoost) and Support Vector Machine (SVM) to lapse behavior modeling. These two advanced algorithms have their merits over other approaches used in the literature such as generalized linear models (i.e., binomial and Poisson models and logistic regression), Classification and Regression Tree (CART) analysis, and the proportional hazards model. Secondly, we adopt economic measures in addition to statistical accuracy in evaluating the performance of different algorithms. Such an adoption better demonstrates how different algorithms may benefit the insurer.

---

[1] There are some papers on the subject of modeling early terminations that do not fit our macro-micro classification on empirical, explanatory studies. They impose specific assumptions on the transition probabilities to early terminations (Buchardt et al., 2015), the early terminations' intensity (Barsotti et al., 2016), or the early termination rates (Loisel and Milhaud, 2011; Milhaud, 2013).

Thirdly, we transform the optimization objective from classification accuracy to economic gains to demonstrate the benefit of integrating modeling with profit maximization. Such an integration can increase life insurers' profitability, improve insurers' customer management through taking preventive measures to reduce lapses, and retain more of the so-called Contractual Service Margin (CSM) in International Financial Reporting Standard (IFRS) 17. It also links us to the literature on churn management and its impact on the customer lifetime value (e.g., Lemmens and Croux, 2006; Lemmens and Gupta, 2017; Neslin et al., 2006).

The results from applying different algorithms to a large dataset consisting of more than six hundred thousand life insurance policies show that XGBoost and SVM outperform CART and logistic regression with respect to statistic accuracy. The results further show that XGBoost is the most robust across training samples.

The advantages of XGBoost and SVM are more apparent with respect to retention gains. The retention gain takes into account the costs of providing incentives to policyholders to reduce their propensities towards lapses, the benefits of retaining policies, and the costs of false alarms. XGBoost and SVM generate much higher retention gains than logistic regression and CART do.

Last but not least, we confirm that economic gains can be further enhanced when the optimization is done on a function linked to the gains rather than on statistic accuracies. The resulted retention gains are 126% of those from applying XGBoost to pursue classification accuracies, and the increase in retention gains remains to be significant under an alternative policyholder retention scheme. An insurer, therefore, should apply robust machine learning algorithms like XGBoost to its economic objective to achieve optimal lapse management.

The organization of the paper is as follows. Section 2 contains explanations about XGBoost and SVM, followed by brief descriptions on CART and logistic regression. In Section 3 we delineate two performance metrics to be used. One is the commonly seen accuracy, i.e., a statistical validation metric, while the other one is an economic metric considering the expected profits and costs of lapse management. We describe the data obtained from a medium-sized life insurer in Section 4. Section 5 displays the comparison results across the four algorithms in terms of the statistical and economic metrics. We explain how to integrate algorithms with the profit maximization goal at the beginning of Section 6, and then compare the results from optimizing profit objectives with those from optimization statistic accuracy. Section 7 summarizes and concludes the paper.

## 2. Binary classification algorithms

The problem that we want to tackle is detecting whether a policyholder will lapse her/his policy or not, i.e., $y_i \in \{0,1\}$. Popular predictive models include logistic regression and CART models. More advanced machine learning models that we introduce in this paper are SVM and XGBoost.

## 2.1. XGBoost

XGBoost is an extension of the gradient boosting introduced by Friedman (2001). The gradient boosting tree is an ensemble method, i.e., multiple weak learners $h$ are combined to become a strong learner $F$ in order to achieve a better predictive performance. The following descriptions are summarized from Friedman (2002).

Given a training sample $\{y_i, \boldsymbol{x}_i\}_1^N$ where $\boldsymbol{x}_i \in \mathbb{R}^n$ and $y_i \in \{0,1\}$, one would like to find a strong learner $F^*(\boldsymbol{x})$ which minimizes a loss function $\Psi(y, F(\boldsymbol{x}))$:

$$F^*(\boldsymbol{x}) = \arg\min_{F(\boldsymbol{x})} E_{y,x}[\Psi(y, F(\boldsymbol{x}))]. \tag{1}$$

The strong learner is an additive expansion of weak learners $h(\boldsymbol{x}, \{R_{lm}\}_1^L, \bar{y}_{lm})$ that will be a $L$-terminal node regression tree in our case:

$$F_M(\boldsymbol{x}) = \sum_{m=0}^{M} \beta_m h(\boldsymbol{x}, \{R_{lm}\}_1^L, \bar{y}_{lm}) = \sum_{m=0}^{M} \sum_{l=1}^{L} \beta_m \bar{y}_{lm} 1(\boldsymbol{x} \in R_{lm}), \tag{2}$$

where $\{R_{lm}\}_1^L$ and $\bar{y}_{lm}$ are the $L$-disjoint regions and the corresponding split points determined by the $m$th regression tree, respectively, and $\beta_m$ are the expansion coefficients. This strong learner is estimated through a stage-wise method that begins with an initial guess $F_0(\boldsymbol{x})$. Then the pseudo-residuals for $m = 1, 2, \ldots, M$ are computed:

$$\tilde{y}_{im} = -\left[\frac{\delta \Psi(y_i, F(\boldsymbol{x}_i))}{\delta F(\boldsymbol{x}_i)}\right]_{F(\boldsymbol{x}) = F_{m-1}(\boldsymbol{x})}. \tag{3}$$

The regions $\{R_{lm}\}_1^L$ are obtained by estimating the $m$th $L$-terminal node regression tree on the sample $\{\tilde{y}_{im}, \boldsymbol{x}_i\}_1^N$. The product $\beta_m \bar{y}_{lm} = \gamma_{lm}$ is set to optimize the loss function $\Psi$:

$$\gamma_{lm} = \arg\min_{\gamma} \sum_{\boldsymbol{x}_i \in R_{lm}} \Psi(y_i, F_{m-1}(\boldsymbol{x}_i) + \gamma). \tag{4}$$

At the final stage, the strong learner is updated,

$$F_m(\boldsymbol{x}) = F_{m-1}(\boldsymbol{x}) + \nu \cdot \gamma_{lm} 1(\boldsymbol{x} \in R_{lm}), \tag{5}$$

where $\nu \in (0,1]$ is a shrinkage parameter that controls how much information is used from the new tree.

The gradient boosting tree method may be summarized as the following algorithm extracted from Friedman (2002).

| | Algorithm 1: Gradient_TreeBoost |
|---|---|
| 1 | $F_0(\mathbf{x}) = \arg\min_{\gamma} \sum_{i=1}^{N} \Psi(y_i, \gamma)$ |
| 2 | For $m = 1$ to $M$ do: |
| 3 | $\quad \tilde{y}_{im} = -\left[\frac{\partial \Psi(y_i, F(\mathbf{x}_i))}{\partial F(\mathbf{x}_i)}\right]_{F(\mathbf{x})=F_{m-1}(\mathbf{x})}, i=1, N$ |
| 4 | $\quad \{R_{lm}\}_1^L = L - \text{terminal node } tree(\{\tilde{y}_{im}, \mathbf{x}_i\}_1^N)$ |
| 5 | $\quad \gamma_{lm} = \arg\min_{\gamma} \sum_{\mathbf{x}_i \in R_{lm}} \Psi(y_i, F_{m-1}(\mathbf{x}_i) + \gamma)$ |
| 6 | $\quad F_m(\mathbf{x}) = F_{m-1}(\mathbf{x}) + \nu \cdot \gamma_{lm} 1(\mathbf{x} \in R_{lm})$ |
| 7 | endFor |

(6)

Inspired by previous general works on statistical learning, many extensions to the gradient boosting tree method have been developed. The stochastic gradient boosting technique (Friedman, 2002) is based on the same principle as the bagging technique (Breiman, 1996). It introduces randomness in the observation: given a random permutation $\pi$ of the integers $\{1, \ldots, N\}$ and $\tilde{N} < N$, the new weak learner tree is estimated on the random subsample $\{\tilde{y}_{\pi(i)m}, x_{\pi(i)}\}_1^{\tilde{N}}$. Another way to inject randomness that has been popularized by Breiman (2001) is randomly selecting a subspace of the explanatory variables. More specifically, given a random permutation $\pi^*$ of integers $\{1, \ldots, n\}$ and $\tilde{n} < n$, the new weak learner tree is estimated on $\{\tilde{y}_{im}, P^*(x)_i\}_1^N$ in which $P^*(x) = \{x_{\pi^*(1)}, \ldots, x_{\pi^*(\tilde{n})}\}$.

To avoid overfitting, some extensions follow the general idea of the ridge regression (Hoerl and Kennard, 1970) and lasso regression (Tibshirani, 1996) and adopt the penalized optimization point of view. Instead of optimizing a loss function $\Psi(y, F(x))$, the problem is modified as the optimization on an "objective" function $O$ that is the sum of a loss function $\Psi$ and a regularization term $\Omega$:

$$O(y, F(x)) = \Psi(y, F(x)) + \Omega(F). \tag{7}$$

Among all the boosting packages that have been developed, the XGBoost system (Chen and Guestrin, 2016) has become the most popular due to its flexibility and computing performances. It has also become the most popular machine learning algorithm in data science challenges such as [Kaggle](#) for structured data. We list the main parameters that need to be tuned, using the package's terminology and the notation of Friedman (2002), as follows.

(i) *nrounds* is the number of trees to grow: $M$;
(ii) *eta* is the shrinkage parameter: $\nu$;
(iii) *gamma* is the regularization parameter which is used in $\Omega$;
(iv) *max_depth* is the number of nodes of a tree: $L$;
(v) *min_child_weight* is the minimal number of observations in a node and $\min_{l,m} \sum_{i=1}^N 1(x_i \in R_{lm})$ should be higher than this value;
(vi) *subsample* is the relative size of the random subsample used in the case of a stochastic gradient boosting: $\tilde{N}/N$;
(vii) *colsample_bytree* is the relative size of the random subspace of explanatory variables selected at each new tree: $\tilde{n}/n$.

Since we are interested in a binary classification in this paper, we use the logistic loss function:

$$\Psi(y, \hat{y}) = \sum_{i=1}^N [y_i \ln(1 + e^{-\hat{y}_i}) + (1 - y_i) \ln(1 + e^{\hat{y}_i})], \tag{8}$$

and the error function as the metric for cross-validation:

$$error(y, \hat{y}) = \frac{\sum_{i=1}^N 1(y_i \neq round(\hat{y}_i))}{N}, \tag{9}$$

where $round(\hat{y}_i) = \begin{cases} 1 \text{ if } \hat{y}_i > 0.5 \\ 0 \text{ if } \hat{y}_i \leq 0.5 \end{cases}$.

The tuning method that we adopt consists of two nested cross-validations. We first perform a grid search on the parameters except *nrounds* with a 2-folds cross-validation (the grid of values is reported in Appendix 9.1). Then we determine the best *nrounds* through a 5-folds cross-validation up to 200 for every possible set of parameters in the grid.

### 2.2. SVM

The theory of SVM was introduced in the 1990's by Boser et al. (1992) and Cortes and Vapnik (1995). It has become a popular algorithm for classification problems and for churn prediction in particular (e.g. Zhao et al., 2005; Xia and Jin, 2008). Its predictive power is rather good compared to other classification algorithms (e.g. Vafeiadis et al., 2015; Wainer, 2016).

The SVM algorithm can be described by geometrical terms. The main idea is to find a hyperplane that separates the observation space into two homogeneous subspaces that is as far apart from each other as possible. This solution is defined as the maximum-margin hyperplane. To deal with misclassifications, a soft margin (i.e., a penalty determined by the user) is imposed upton the SVM. Another way to deal with classification errors is to project the data to a higher-dimensional space through a kernel function. A more complete geometrical description of SVM can be found in Noble (2006).

In the following, we adopt a formula-based description of the SVM by using the notation of Hsu et al. (2003). Given a training sample $\{y_i, x_i\}_1^N$ in which $x_i \in \mathbb{R}^n$ and $y_i \in \{+1, -1\}$, the SVM algorithm is the solution of the following optimization problem:

$$\min_{\omega,b,\xi} \frac{1}{2} \omega^T \omega + C \sum_{i=1}^N \xi_i, \tag{10}$$

with the constraint

$$y_i(\omega^T \phi(x_i) + b) \geq 1 - \xi_i, \ \xi_i \geq 0. \tag{11}$$

The separating hyperplane is determined by the orthogonal vector $\omega$ and constant $b$. The soft margin penalty cost is denoted as $C$. The data may be projected to a higher dimension space by the function $\phi$, and the underlying kernel function is defined by $K(x_i, x_j) = \phi(x_i)^T \phi(x_j)$.

In our case we choose to consider the radial basis function kernel (also called RBF kernel) that is the most commonly used in practice and determined by

$$K(x_i, x_j) = \exp(-\gamma \|x_i - x_j\|^2), \tag{12}$$

with $\gamma > 0$ being the kernel parameter.

Then we use the "e1071" R package (Meyer et al., 2015) to implement the SVM algorithm. To tune the SVM parameters $(C, \gamma)$, we perform a grid search on a 2-folds cross-validation and adopt the misclassification error function as the validation metric. The grid of values is reported in Appendix 9.2.

### 2.3. CART

CART was first introduced by Breiman (1984). The underlying idea is straight forward: defining a class by following a list of decision rules on the explanatory variables. To determine these rules, the data space is iteratively separated by binary split into two disjointed subspaces. At each step or node of this top-down construction, the explanatory variable and the dividing point are chosen to minimize the Gini impurity of the node.

More specifically, given a node $l$ of $N_l$ observations of response $y_i \in \{0,1\}$ with $i \in l$, the proportion of observations in the node is defined by $p_l = \frac{1}{N_l} \sum_{i \in l} y_i$. Then use an algorithm to partition the parent node into two nodes $l_L$ and $l_R$ by maximizing

$$I_G(l) - [I_G(l_L) + I_G(l_R)], \tag{13}$$

where $I_G$ is the Gini impurity of the node and computed by

$$I_G(l) = N_l p_l (1 - p_l). \tag{14}$$

This construction is applied up to obtaining a node for every observation point. The tree obtained is thus designated as the saturated model. Although fitting the response on the training sample perfectly, it generally leads to low predictive performance when applied to new samples. Hence the tree needs to be pruned, i.e., the number of final nodes needs to be reduced to increase its predictive power.

Many criteria can be used to prune the tree, e.g., the minimum number of observations in a final node. We choose $L$, the number of terminal nodes, that minimizes the misclassification error:

$$error(y, \hat{y}) = \frac{\sum_{i=1}^{N} \mathbf{1}(y_i \neq \hat{y}_i)}{N}. \tag{15}$$

$L$ is estimated by a 10-folds cross-validation methodology. We use the "rpart" R package (Therneau et al., 2018) to implement CART.

### 2.4. Logistic regression

The logistic regression is a special case of the generalized linear models (Nelder and Wedderburn, 1972) obtained with the Bernoulli distribution. The goal is to model the probability of a binary event such as the lapse probability $p_i$ of the policyholder $i$. Given a training sample $\{y_i, x_i\}_1^N$ in which $x \in \mathbb{R}^n$ and $y_i \in \{0,1\}$, the regression model is specified as:

$$\ln \frac{p_i}{1-p_i} = \beta_0 + x_i^T \boldsymbol{\beta}. \tag{16}$$

The parameters $(\beta_0, \boldsymbol{\beta}) \in \mathbb{R} \times \mathbb{R}^n$ can be estimated by the maximum-likelihood method:

$$\mathcal{L} = \prod_{i=1}^{N} \left( \frac{e^{x_i^T \beta}}{1+e^{x_i^T \beta}} \right)^{y_i} \left( \frac{1}{1+e^{x_i^T \beta}} \right)^{1-y_i}. \tag{17}$$

When applying the estimated logistic regression model to a classification problem, it doesn't directly lead to labeled responses but to estimated probabilities. To determine the forecasted class, we chose the common threshold of 0.5, i.e.,

$$\hat{y}_i^* = \begin{cases} 1 \text{ if } \hat{y}_i > 0.5; \\ 0 \text{ if } \hat{y}_i \leq 0.5. \end{cases} \tag{18}$$

### 3. Validation metrics

For each policy, the observed lapse $y_i$ and the forecasted lapse $\hat{y}_i$ are binary variables: $(y_i, \hat{y}_i) \in \{0,1\}^2$. The four different outputs of a binary classification model are named true positive (1,1), true negative (0,0), false positive (0,1) and false negative (1,0) while the number of each case is usually laid out in the so-called confusion matrix. Denote $N(j, k)$ as the numbers of the confusion matrix in which $j \in \{0,1\}$ stands for the observed lapse indicator and $k \in \{0,1\}$ the predicted lapse indicator. Given a set of response variables $\{y_i, \hat{y}_i\}_1^N$, we estimate $N(j,k)$ as:

$$N(j,k) = \sum_{i=1}^{N} \mathbf{1}(y_i = j, \hat{y}_i = k). \tag{19}$$

### 3.1. Statistical metric

Based on the confusion matrix, different metrics can be developed. We first focus on the accuracy metric, the ratio of correctly classified predictions over the total number of predictions:

$$accuracy(y, \hat{y}) = \frac{N(1,1) + N(0,0)}{N} = 1 - error(y, \hat{y}). \tag{20}$$

### 3.2. Economic metric

Although we adopt mathematic algorithms to predict lapses, the risk is an economic issue after all. We thus would like to analyze and compare the classification algorithms by an economic metric. More specifically, we will estimate the impacts of different classification results on the expected profits from policies, also called customer lifetime values. In order to do so, we plan to adopt an economic model inspired by Neslin et al. (2006) and Gupta et al. (2006).

Suppose that policy $i$ stays $\Theta_i$ years in the portfolio ($\Theta_i \in \mathbf{N}$). The profitability ratio at time $t$ can be represented by $p_{i,t}$ and the face amount by $F_{i,t}$. The lifetime value for policy $i$ is computed as:

$$CLV_i = \sum_{t=0}^{\Theta_i} \frac{p_{i,t} F_{i,t}}{(1+d_t)^t}, \tag{21}$$

where $d_t$ is the discount rate.

Assuming a deterministic time horizon T ($T \in \mathbf{N}$), we define the $(T+1)$-dimensional real vectors $\boldsymbol{p}_i, \boldsymbol{F}_i, \boldsymbol{r}_i$, and $\boldsymbol{d}$ for profitability ratios, face amounts, retention probabilities, and interest rates respectively. Given the four vectors, the customer lifetime value is

$$CLV_i(\boldsymbol{p}_i, \boldsymbol{F}_i, \boldsymbol{r}_i, \boldsymbol{d}) = \sum_{t=0}^{T} \frac{p_{i,t} F_{i,t} r_{i,t}}{(1+d_t)^t}. \tag{22}$$

The lapse management strategy is modelled by the offer of an incentive $\boldsymbol{\delta}_i \in \mathbb{R}^{T+1}$ to policyholder $i$ who is contacted with a cost $c$. The incentive is accepted with the probability $\gamma_i$, and the acceptance will change the vector of the probabilities of staying in the portfolio from $\boldsymbol{r}_i$ to $\boldsymbol{r}_i^* \in \mathbb{R}^{T+1}$. We further make the following simplifying assumptions:

(i) $\boldsymbol{p}_i$ are the same for all policies and denoted as $\boldsymbol{p}$ hereafter;
(ii) $\boldsymbol{\delta}_i$ are the same for all contacted policies and denoted as $\boldsymbol{\delta}$ hereafter;
(iii) $p_{i,t}, F_{i,t}$ and $d_t$ remain constant across time;
(iv) $\boldsymbol{r}_i$ equals to $\boldsymbol{r}_{lapse}$ or $\boldsymbol{r}_{stay}$ in which $\boldsymbol{r}_{stay} = (1,1,\dots,1)$ and $\boldsymbol{r}_{lapse}$ is estimated on the dataset and will be given in Section 5.2;
(v) if $\boldsymbol{r}_i = \boldsymbol{r}_{stay}$, the incentive is accepted with probability $\gamma_i = 1$ and $\boldsymbol{r}_i^* = \boldsymbol{r}_{stay}$;

(vi) if $r_i = r_{lapse}$, the incentive is accepted with probability $\gamma_i = \gamma$ and $r_i^* = r_{stay}$.[2] Policyholders who reject the offers (probability = 1- $\gamma$) will lapse their policies, i.e. $r_i^* = r_{lapse}$.

The application of a segmentation algorithm to the tested samples produces two confusion matrices: one with respect to number of policies while the other in term of face amount. For the latter matrix, we denote $F(j,k)$ as the coefficients of the matrix with regard to face amount, where $j$ stands for the indicator of the policyholder's lapse in real life, $k$ the indicator by the algorithm's prediction, and $(j,k) \in \{0,1\}^2$. More specifically,

$$F(j,k) = \sum_{i=1}^{N} F_i \cdot \mathbf{1}(y_i = j, \hat{y}_i = k), \tag{23}$$

while $N$ is defined in Equation (19).

We define the reference portfolio value (RPV) as the customer lifetime value of all policies if no customer relationship management about lapses are carried out to be:

$$RPV = CLV(\boldsymbol{p}, F(0,0) + F(0,1), \boldsymbol{r}_{stay}, \boldsymbol{d}) \\ + CLV(\boldsymbol{p}, F(1,0) + F(1,1), \boldsymbol{r}_{lapse}, \boldsymbol{d}). \tag{24}$$

Given a segmentation algorithm, we compute the lapse managed portfolio value (LMPV) by

$$LMPV(\boldsymbol{\delta}, \gamma, c) = CLV(\boldsymbol{p}, F(0,0), \boldsymbol{r}_{stay}, \boldsymbol{d}) + CLV(\boldsymbol{p}, F(1,0) + (1-\gamma)F(1,1), \boldsymbol{r}_{lapse}, \boldsymbol{d}) \\ + CLV(\boldsymbol{p} - \boldsymbol{\delta}, F(0,1) + \gamma F(1,1), \boldsymbol{r}_{stay}, \boldsymbol{d}) - c(N(0,1) + N(1,1)). \tag{25}$$

Then we define the economic metric of the algorithm as the retention gain:

$$RG(\boldsymbol{\delta}, \gamma, c) = LMPV(\boldsymbol{\delta}, \gamma, c) - RPV, \tag{26}$$

that can be simplified as

$$\gamma \left[ CLV(\boldsymbol{p} - \boldsymbol{\delta}, F(1,1), \boldsymbol{r}_{stay}, \boldsymbol{d}) - CLV(\boldsymbol{p}, F(1,1), \boldsymbol{r}_{lapse}, \boldsymbol{d}) \right] \\ - CLV(\boldsymbol{\delta}, F(0,1), \boldsymbol{r}_{stay}, \boldsymbol{d}) - c(N(0,1) + N(1,1)). \tag{27}$$

## 4. Data

Our data come from a medium-size life insurance company in Taiwan that had total assets over 15 billion US dollars at the end of 2013. The data contain 629,331 life insurance policies sold during the period from 1998 to 2013. The data-providing insurer tracked changes in the statuses of policies including death and lapse. The last tracking date is 8/31/2013. 243,152 policies out of all samples were lapsed, and 5,486 insureds died during the sampling period.

We specify several variables based on the literature and the data provided by the insurer as input to the algorithms of Section 2. Firstly we are able to identify from the data the age, gender, and occupation of an insured at the time when the policy was issued. Female is designated as 1 while male 0 for the dummy variable Gender. Then we designate the dummy variable Occupation as 1 for the occupations that the insurers in Taiwan would undertake extra screening/underwriting. The data also record whether the insured is required to have a physical examination when purchasing life insurance and how many non-life policies (health and long-term care) a person are listed as the insured (since a person may purchase multiple policies).

---

[2] These simplifications assume that the profitability ratio, the incentive, and the probability to accept the incentive is the same across policyholders, respectively. Upon the availability of data, we may compute an expected profitability ratio for each policy. The incentive offered to each policyholder can then be set as a function of the policy's profitability. The probability of accepting the offer can also be a function of the incentive, but such a function is difficut to estimate in practice. Face amount may be variable for some products, which increases the difficulty in estimating the expected profitability ratio. The retention probabilities may change with time, and this calls for a dynamic model of lapse propensities.

The data also contain the inception date and face amount of each policy. There are three types of policies. The most popular type is traditional policies like term life, whole life, and endowment. Investment-linked and interest-adjustable types of products appeared in 2000s. We also able to identify whether a policy is a single-premium one or not. There are three cases with regard to participation. It was not until 2004 that insurers were allowed to sell non-participating policies. The policies sold by the end of 2003 are thus designated as Mandotory Participating. Starting from 2004, policies may be classified into participating and non-participating. Most policies sold in Taiwan are dominated in New Taiwan Dollar (NTD) ; there are some policies dominated in other currencies.

We further set up two nominal variables. Firstly, we categorize distribution channels as Tied Agents (denoted by TA), Direct Marketing (DM), and Banks (BK)[3]. Secondly, premium paying methods are classified into three ways: collected by the personnel of the insurer (denoted as Insurer), automatic transfers from banks or payments by credit cards (B&C),[4] and going to the post office or convenient stores in person (P&C).

Table 1 presents the descriptive statistics of the above explanatory variables. The average age of the sampled insureds is 28 and the standard deviation of the insureds' age is 17. The minimum, medium, and maximum age is 0, 27, and 80, respectively. The samples consist of relatively equivalent portions of male and female insureds. About 20% of the insureds work in riskier occupations that call for extra underwriting. Most insureds (over 96%) were not required to go through physical examination in purchasing life insurance. Many insureds are associated with multiple non-life policies so that the average number of non-life policies a person are listed as the insured is 1.2. There is a person who is listed as the insured for 33 non-life policies.

The mean and medium of policy inception dates are in the second quarter of 2005, and the standard deviation around this quarter is almost 5 years. The face amount of the sampled policies has an average of 17,165 US dollars[5] with big variations: the largest policy reaches 2 million dollars, the smallest one is only 333 dollars,[6] and the standard deviation is about twenty-eight thousand dollars. Around 3% of the samples are single-premium policies. 46.6% of samples are mandatory-participating policies while 37.2% are non-participating ones. Almost all policies are traditional types of products ; interest-adjustable and investment-linked types of products are merely 3% of our samples. 88% of policies are dominated in NTD.

---

[3] Few policies are also sold by independant agents, brokers that we gather in the same category.

[4] Paying premiums by automatic transfers from bank accounts or by recurring payments of credit cards is indifferent to policyholders. We thus regard these two automatic/recurring payment methods as one.

[5] The exchange rate used in the paper is 30 NTD/1 USD.

[6] This policy is a whole life insurance with a one-year old insured and the death benefit of ten thousand NTD (a little over three hundred USD). There are other small policies with death benefits smaller than three thousand USD. These policies constitute less than one percent of our samples.

Table 1 also shows that selling life insurance through tied agents is the major way (94%) of this insurer while the sampled policies sold through direct marketing are smaller than 3%. It further shows that the most popular way of paying premiums is through automatic/recurring transfers from bank accounts or credit cards (71%). Since post offices and convenient stores providing money transferring services are conveniently around, about 10% of our samples have premiums paid in places like these.

Table 1: Descriptive Statistics of Explanatory Variables.

| Variables | Percentage |
|---|---|
| Gender | |
| *Female* | 48% |
| *Male* | 52% |
| Occupation | |
| *Tier one* | 80.5% |
| *Requiring extra screening* | 19.5% |
| Physical Examination | |
| *Exempted* | 96.4% |
| *Required* | 3.6% |
| Distribution Channel | |
| *TA* | 93.9% |
| *BK* | 3.4% |
| *DM* | 2.4% |
| *Others[7]* | 0.3% |
| Premium payment | |
| *Single premium* | 3.1% |
| *Non single premium* | 96.9% |
| Premium Paying Method | |
| *Insurer* | 18.8% |
| *B&C* | 70.8% |
| *P&C* | 10.4% |
| Participation | |
| *Non-participating* | 37.2% |
| *Participating* | 16.2% |
| *Mandatory participating* | 46.6% |
| Product Type | |
| *Interest-Adjustable* | 1.7% |
| *Investment-Linked* | 1.2% |
| *Traditional* | 97.1% |
| Currency Domination | |
| *NTD* | 88.1% |
| *Others* | 11.9% |

---

[7] Few were sold through independent agents or brokers.

|  | Mean | Medium | Standard Deviation | Minimum | Maximum |
|---|---|---|---|---|---|
| Age | 28.3 | 27 | 16.8 | 0 | 80 |
| # of non-life policies | 1.2 | 0 | 2 | 0 | 33 |
| Inception date | 06/06/2005 | 21/04/2005 | 4,8 (years) | 01/01/1998 | 31/07/2013 |
| Face Amounts (in USD) | 17,165 | 10,000 | 28,050 | 333 | 2,000,000 |

## 5. Results with respect to statistical and economic metrics

Our focus is on the predictive performance of different algorithms. We thus conduct out-of-sample tests using the following procedure. First, we randomly split the dataset D into 10 subsamples $\{D_1, ..., D_{10}\}$ of equal size and then train an algorithm on $D_k$, k ∈{1,…,10}. The estimated model is subsequently applied to the other subsamples to obtain forecasts $\hat{y}$ of lapses. In the last step, we compare these predictions with the observed lapses $y$ by the validation metric $\rho(y,\hat{y})$ to measure the predictive performance of the algorithm. This procedure enables us to make sure that every observation is used, at some point of an algorithm, as both training and testing samples. It is similar to the k-fold cross-validation technique in which the training subsample is composed of $D - D_k$ and the testing subsample is set to $D_k$. We use the k-fold cross-validation to tune parameters in training some of the algorithms.

### 5.1. Results with respect to the statistical metric

The mean accuracy computed using the above cross-validation procedure is displayed in the Table 2 and Figure 1 for each binary classification algorithm. As expected, the more sophisticated the model is, the more accurate the predictions will be. XGBoost ranks number one, followed by SVM, CART, and logistic regression (LR). XGBoost surpasses logistic regression by 2.24% on average, which represents a significant improvement of 12,684 correctly classified policies. Moreover, the smallest standard deviation of accuracy of the XGBoost, 0.03%, indicates that XGBoost is less prone to sample selection. This is visible in the box plot of Figure 1.

Table 2 – Cross-Validated Statistic Accuracies

|  | LR | CART | SVM | XGB |
|---|---|---|---|---|
| Mean Accuracy | 76.64% | 77.15% | 77.82% | 78.88% |
| Standard Deviation | 0.07% | 0.10% | 0.08% | 0.03% |

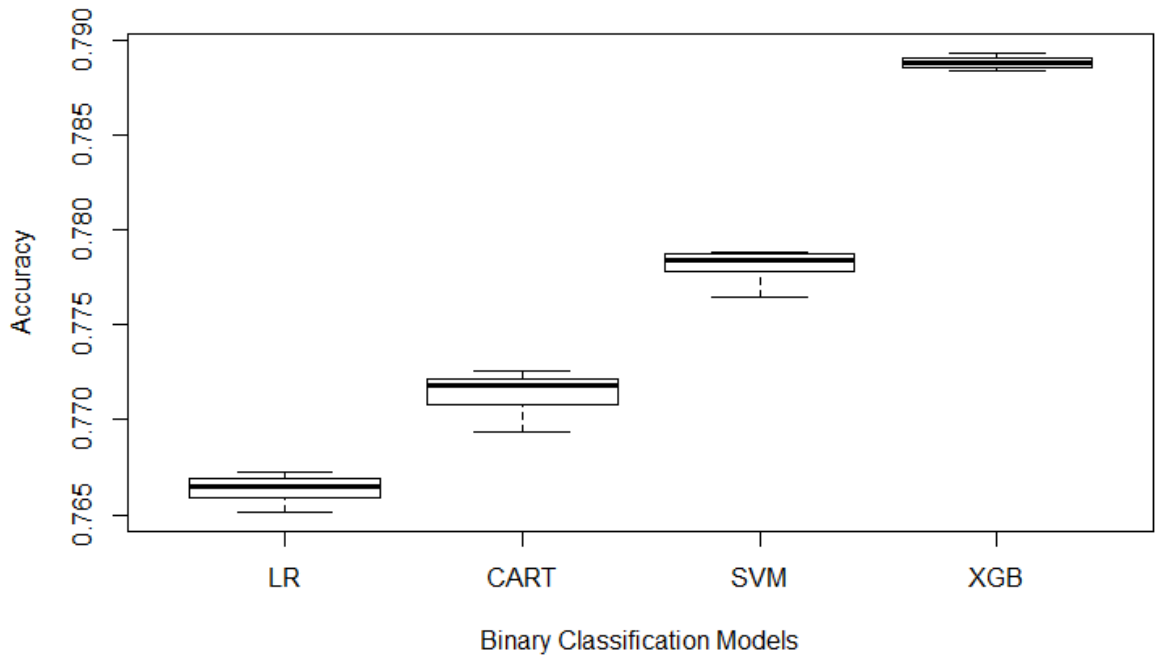

Figure 1: Box Plot of Statistic Accuracies

Looking at the entire confusion matrices in Tables 3a to 3d, we find that CART predicts the most lapses (191,869 = 51,241 + 140,628) from which it identifies the most lapses correctly (140,628) but also signals the most false alarms (51,241). SVM predicts the most stays (398,597 = 310,258 + 88,339) in which it identifies the most stays correctly (310,258) while produces many false security cases (88,339). XGBoost is rather robust on the other hand. It is ranked the second in terms of all aspects: correctly identifying lapses (137,660), correctly identifying stays (309,111), not producing false alarms (38,450), and not producing false securities (81,177).

Table 3a: Average Confusion Matrix of XGB

|  |  | Predicted | |
|---|---|---|---|
|  |  | Stay | Lapse |
| **Actual** | Stay | 309,111 | 38,450 |
|  | Lapse | 81,177 | 137,660 |

Table 3b: Average Confusion Matrix of SVM

|  |  | Predicted | |
|---|---|---|---|
|  |  | Stay | Lapse |
| **Actual** | Stay | 310,258 | 37,303 |
|  | Lapse | 88,339 | 130,498 |

Table 3c: Average Confusion Matrix of CART

|  |  | Predicted | |
|---|---|---|---|
|  |  | Stay | Lapse |
| **Actual** | Stay | 296,320 | 51,241 |
|  | Lapse | 78,209 | 140,628 |

Table 3d: Average Confusion Matrix of LR

|  |  | Predicted | |
|---|---|---|---|
|  |  | Stay | Lapse |
| **Actual** | Stay | 304,025 | 43,537 |
|  | Lapse | 88,775 | 130,062 |

## 5.2. Results with respect to the economic metric

To evaluate the algorithms by the economic metric, we first need to specify the parameters of the cash flows model. Since no data is available for us to estimate these parameters, we have to make assumptions. We had conducted sensitivity analyses and confirmed that the comparison results remain the same in general.

The time horizon $T$ is set to 12 years according to the length of the sampling period. We estimate the retention probability vector $r_{lapse}$ from the dataset and obtain:

| Year t | 0 | 1 | 2 | 3 | 4 | 5 | 6 | 7 | 8 | 9 | 10 | 11 | 12 |
|---|---|---|---|---|---|---|---|---|---|---|---|---|---|
| Retention probability | 0.96 | 0.87 | 0.67 | 0.37 | 0.27 | 0.21 | 0.15 | 0.12 | 0.1 | 0.08 | 0.06 | 0.05 | 0.04 |

Other parameters are set as follows:
- the profitability ratio $p = 0.5\%$;
- the discount rate $d = 2\%$;
- the cost to contact a policyholder $c = 10$ USD.

We propose two different incentive strategies: an aggressive one and a moderate one. The incentive vectors are defined as below:

| Year t | 0 | 1 | 2 | 3 | 4 | 5 | 6 | 7 | 8 | 9 | 10 | 11 | 12 |
|---|---|---|---|---|---|---|---|---|---|---|---|---|---|
| **Incentive 1** | 0% | 0% | 0.030% | 0.030% | 0.060% | 0.060% | 0.090% | 0.090% | 0.120% | 0.120% | 0.150% | 0.150% | 0.180% |
| **Incentive 2** | 0% | 0% | 0.015% | 0.015% | 0.030% | 0.030% | 0.045% | 0.045% | 0.060% | 0.060% | 0.060% | 0.060% | 0.060% |

We further assume that the probabilities of accepting the incentives for a would-lapse policyholder are $\gamma_1 = 20\%$ and $\gamma_2 = 10\%$ respectively.

The results from comparing different classification algorithms by the economic metric with the aggressive incentive strategy are displayed in Table 4 and Figure 2. The winner looks to be XGBoost: it has the highest retention gain with the smallest standard deviation across subsampling. Figure 2 further illustrates that XGBoost and SVM lead to similar retention gain compared to logistic regression and CART.

Notice that the differences across the algorithms are wider in terms of the economic metric than the statistical metric. The accuracies of the models are between 76.64% and 78.88%, which means an improvement ratio of 2.9%. The retention gains, on the other hand, range from 2.7 and 5.2 million USD, indicating an enhancement of 96%. Therefore, choosing a good algorithm is more important in terms of economic reality (dollar amount) than by statistical accuracy.

It appears that CART produces the lowest retention gain: $2,680,012. This is mostly because CART has the highest false alarm rate (cf. Table 3c) which means offering the incentive to many policyholders who have no intention to lapse their policies. Furthermore, CART leads to the highest contacting cost since it predicts the highest lapses. The profits are thus reduced.

Table 4: Cross-Validated Retention Gains with the Aggressive Strategy

|  | LR | CART | SVM | XGB |
|---|---|---|---|---|
| Mean Retention Gain | 4,046,602 | 2,680,012 | 5,028,737 | 5,243,913 |
| Standard Deviation | 133,993 | 209,220 | 139,102 | 115,415 |

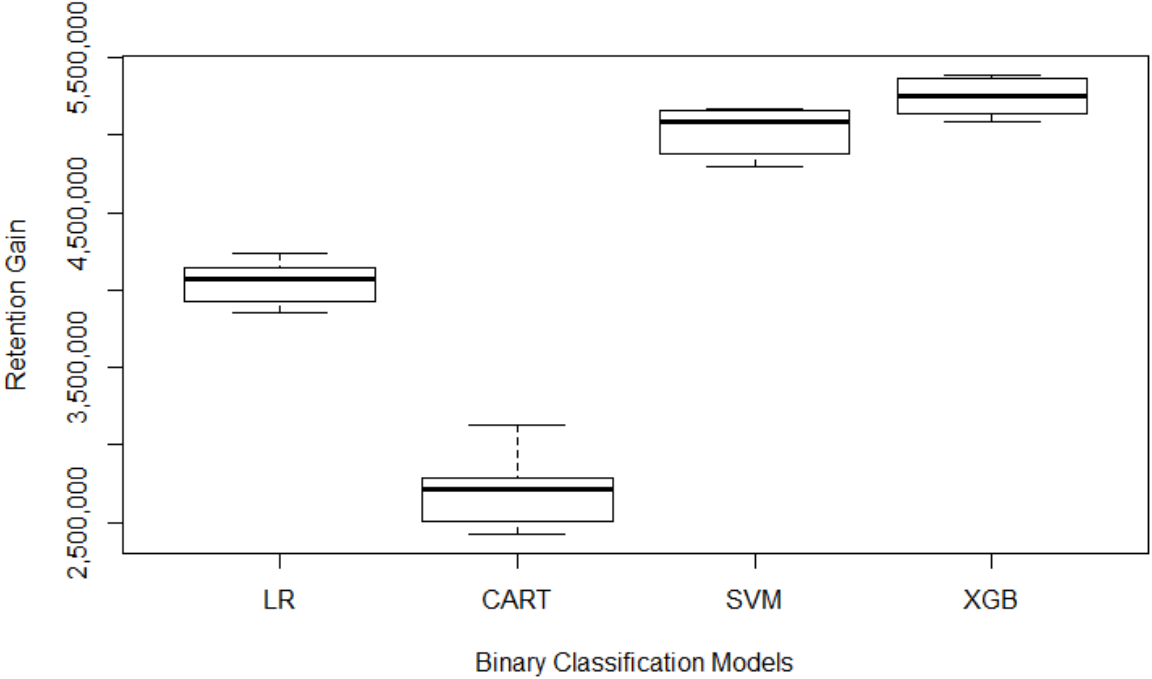

Figure 2: Box Plot of Retention Gains with the Aggressive Strategy

Then we look at algorithms' performances when the incentive strategy is moderate and leads to lower acceptance probabilities. The results are displayed in the Table 5 and the Figure 3. We first notice XGB and SVM remains to be ranked No. 1 and No. 2, respectively. Next we observe that the improvement ratio of the best algorithm over the worst is smaller but remains to be significant (56%). Thirdly, retention gains are significantly lower with the moderate incentive strategy. For instant, XGB achieves a gain of 5.2 million dollars with the aggressive incentive strategy but the gain reduces to 3.3 million dollars when incentives offered to policyholders are moderate. Under our assumptions, the company should rather set the aggressive incentive strategy up to optimize her gains. However, in practice, one would need a more complete sensitivity study on the incentive to be offered and the corresponding acceptance probability to fully optimize the lapse management.

Table 5: Cross-Validated Retention Gains with the Moderate Strategy

|  | LR | CART | SVM | XGB |
|---|---|---|---|---|
| Mean Marketing Gain | 2,618,396 | 2,085,599 | 3,113,900 | 3,261,029 |
| Standard Deviation | 63,693 | 85,184 | 54,169 | 45,928 |

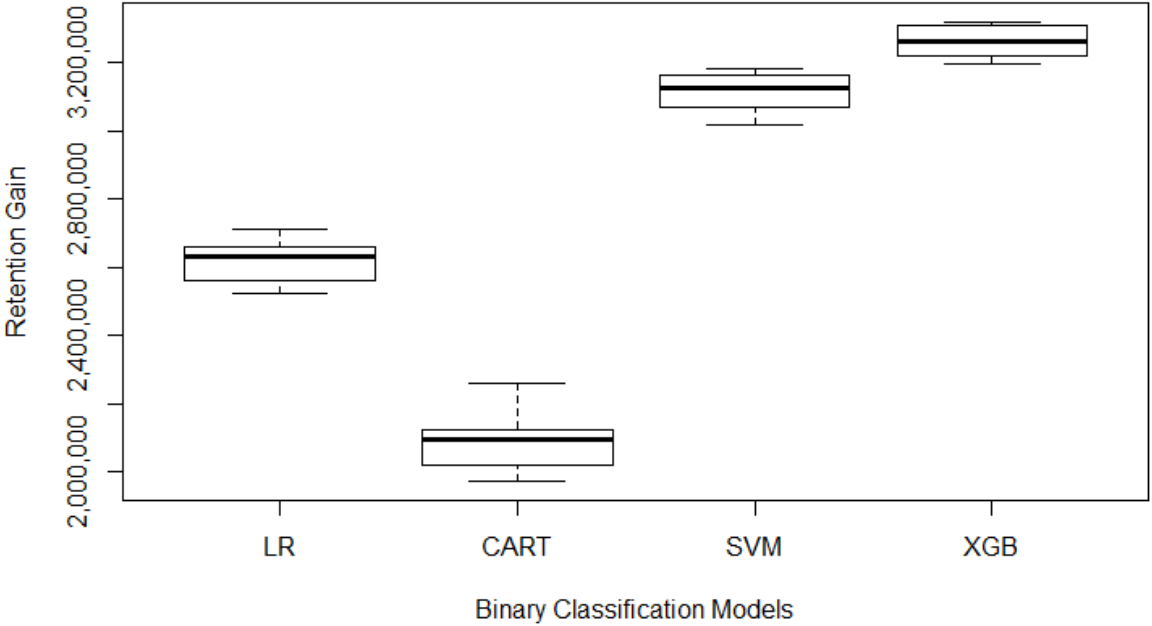

Figure 3: Box Plot of Retention Gains with the Moderate Strategy

In summary, XGB and SVM consistently perform better than CART and LR no matter which performance index, statistical accuracy or retention gains with alternative incentive strategies, is used. The drawbacks of XGB and SVM relative to CART and LR that we may think of are not related to performance. For instance, XGB and SVM are less transparent, more complex, demanding more computing power, and more difficult to be comprehended by inexperienced persons than CART and LR.

## 6. Optimization on profitability instead of classification

It is obvious that insurers would not seek to optimize the classification accuracy but focus on economic gains resulted from the classification algorithms when forming a lapse management strategy. When our aim is to maximize the profitability of the lapse management strategy, binary classifications might be unsuitable since they are not designed to meet such a need. Ascarza et al. (2018) emphasize the difference between the at-risk population (e.g., customers with high churn probabilities) and the targeted population (e.g., customers that the company should focus her retention campaign on in order to optimize her profits) from an economic point of view. Along this line of churn literature, Lemmens and Gupta (2017) modify the usual loss function into a profit-based function to optimize economic gains. They obtain a significantly increase in the expected profit of a retention campaign. Learning from the churn literature, we transform the above classificaton problem into a regression question in this section.

### 6.1. Methodology

Let the new response variable $z_i^{R_j}$ represents the retention gain or loss resulting from proposing the incentive $j \in \{1,2\}$ (cf. Section 5.2) to policyholder $i$. More specifically, we define $z_i^{R_j}$ as

$$z_i^{R_j} = \begin{cases} -CLV(\boldsymbol{\delta}_j, F_i, \hat{\boldsymbol{r}}, \boldsymbol{i}) - c & \text{if } y_i = 0, \\ \gamma_j \cdot [CLV(\boldsymbol{p} - \boldsymbol{\delta}_j, F_i, \hat{\boldsymbol{r}}, \boldsymbol{i}) - CLV(\boldsymbol{p}, F_i, \boldsymbol{r}_{lapse}, \boldsymbol{i})] - c & \text{if } y_i = 1; \end{cases} \quad (29)$$

Then we may apply the XGBoost algorithm to $\{z_i^{R_j}, \boldsymbol{x}_i\}_1^N$ and use the mean squared error as the loss function

$$\Psi\left(z^{R_J}, \widehat{z^{R_J}}\right) = \frac{1}{N}\sum_{i=1}^N \left[z_i^{R_j} - \widehat{z^{R_J}}_i\right]^2, \quad (30)$$

and as the metric for cross-validation.

In the last step, lapse $\hat{y}_i$ is forecasted if the estimated gain is positive:

$$\hat{y}_i = \begin{cases} 1 & \text{if } \widehat{z^{R_J}}_i > 0 \\ 0 & \text{if } \widehat{z^{R_J}}_i \leq 0 \end{cases}, \quad (31)$$

By this way we can apply the same metrics described in previous sections. Here $\hat{y}_i$ is better to be understood as the estimation of the profitability about offering an incentive to the policyholder $i$ rather than the forecast on the policyholder's lapse.

The two new classifications are denoted as XGB_R1 and XGB_R2, respectively, for applying XGBoost to $z^{R_1}$ and $z^{R_2}$. The tuning method that we apply to estimating the parameters is described in Appendix 9.3.

### 6.2. Results

Table 6 and Figure 4 display the prediction accuracies. Table 6 shows that XGB_R1 and XGB_R2 produce relatively low mean accuracy of respectively 76.7% and 75.7% While XGB_R2 is clearly the worst model in term of accuracy, XGB_R1 generates similar results to the logistic regression which is the worst binary classification model regarding the accuracy measure. These seemingly unsatisfied results are understandable since both XGB_R1 and XGB_R2 are not designed to predict whether a policy would be lapsed or not. What they aim for are economic gains.

Table 6: Cross-Validated Accuracy

|  | LR | CART | SVM | XGB | XGB_R1 | XGB_R2 |
|---|---|---|---|---|---|---|
| Mean Accuracy | 76.64% | 77.15% | 77.82% | 78.88% | 76.67% | 75.71% |
| Standard Deviation | 0.07% | 0.10% | 0.08% | 0.03% | 0.07% | 0.06% |

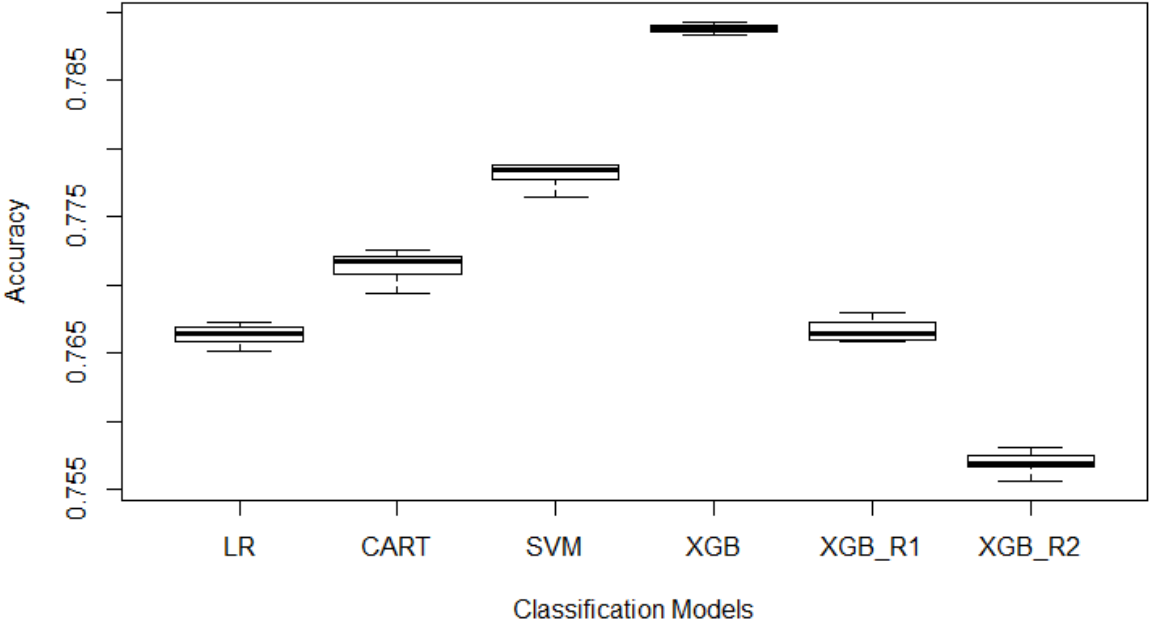

Figure 4: Box Plot of Cross-Validated Accuracy

The numbers in Table 7a and 7b tell us more about why XGB_R1 and XGB_R2 performs badly in statistical accuracy. They result in the smallest correct identifications on lapses (resp. 104,889 and 99,432) and produce the most false-sense-of-security (resp. 113,948 and 119,405). However, we will see very soon that XGB_R1 and XGB_R2 stand out when we switch focus to retention gain.

Table 7a: Average Confusion Matrix of XGB_R1

|  |  | Predicted | |
|---|---|---|---|
|  |  | Stay | Lapse |
| **Actual** | Stay | 329,357 | 18,204 |
|  | Lapse | 113,948 | 104,889 |

Table 7b – Average Confusion Matrix of XGB_R2

|  |  | Predicted | |
|---|---|---|---|
|  |  | Stay | Lapse |
| **Actual** | Stay | 329,413 | 18,149 |
|  | Lapse | 119,405 | 99,432 |

Table 8 and Figure 5 show that XGB_R1 generates a significantly larger average retention gain with the aggressive incentive strategy ($6,586,357) than other algorithms as well as a significantly lower standard deviation ($53,460). The increase in retention gain is 26% (1.3 million USD) higher than that generated by XGB (the second-best algorithm) and 146% (3.9 million USD) better than that produced by CART. Looking back to Table 7a, we see that XGB_R1 leads to reduce the number of false alarms (18,204) in optimizing the retention gain, even if this also reduces the correct detection (104,889). The good results of XGB_R1 in achieving retention gain demonstrate the benefit of integrating the algorithm with the goal to be achieved. The objective function for XGB_R1 to minimize, Equation (30), is about predicting retention gains. XGB_R1 therefore would naturally perform the best when compared with other algorithms optimizing other objectives (such as classification accuracies).

Table 8: Cross-Validated Retention Gains with the Aggressive Strategy

|  | LR | CART | SVM | XGB | XGB_R1 |
|---|---|---|---|---|---|
| Mean Retention Gain | 4,046,602 | 2,680,012 | 5,028,737 | 5,243,913 | 6,586,357 |
| Standard Deviation | 133,993 | 209,220 | 139,102 | 115,415 | 53,460 |

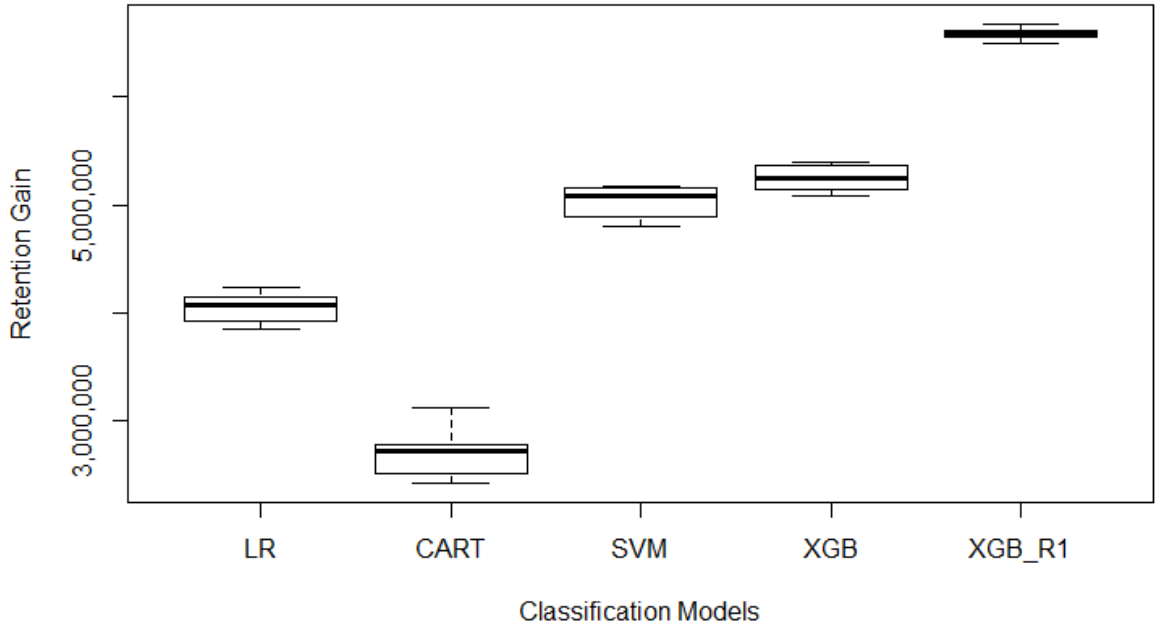

Figure 5: Box Plot of Retention Gains the Aggressive Strategy

We expect that the benefit of integrating the algorithm with the goal is robust across incentive strategies. This is confirmed by the results in Table 9 and Figure 6. XGB_R2 generates retention gain of 3.9 million dollars that is nearly 600 thousand dollars more than that achieved by the second place XGB. The increase in retention gains is 18%. The increases with respect to the commonly seen LR and CART reach 47% and 85%.

Table 9: Cross-Validated Retention Gains the Moderate Strategy

|  | LR | CART | SVM | XGB | XGB_R2 |
|---|---|---|---|---|---|
| Mean Marketing Gain | 2,618,396 | 2,085,599 | 3,113,900 | 3,261,029 | 3,852,782 |
| Standard Deviation | 63,693 | 85,184 | 54,169 | 45,928 | 39,163 |

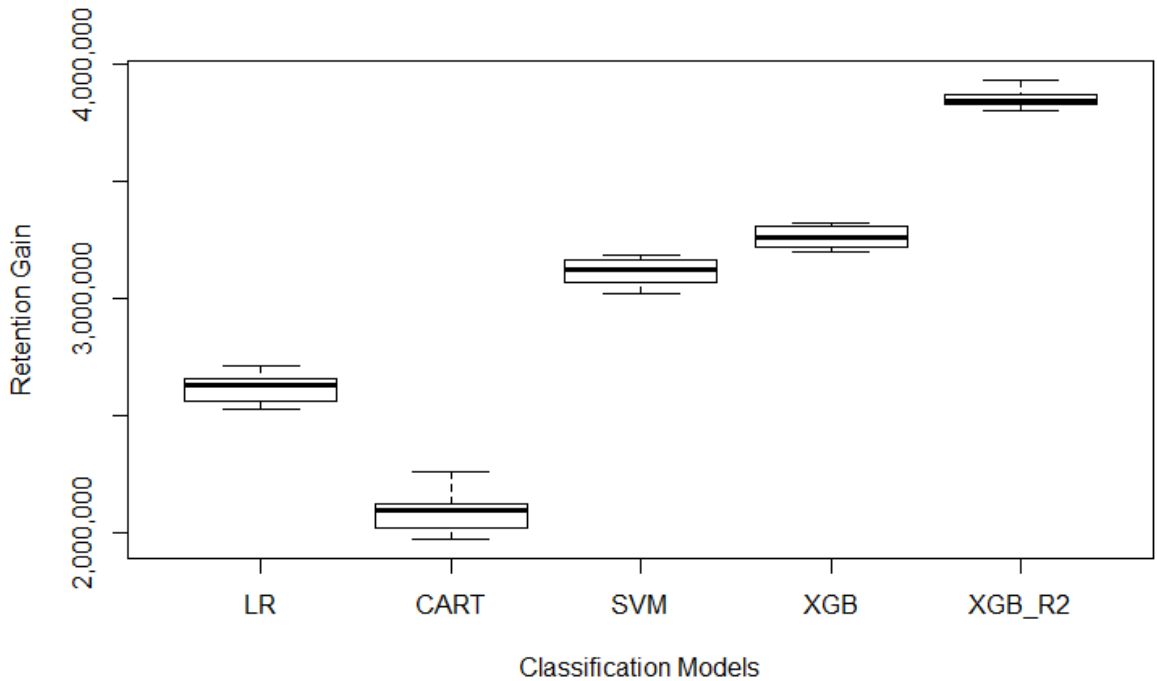

Figure 6: Box Plot of Retention Gains the Moderate Strategy

The results in this section demonstrate the benefit of having a specific objective. If senior managers of an insurer are able to specify an objective to be optimized (e.g., maximizing retention gain), the staff should apply an advanced algorithm like XBG directly to such an objective to achieve the optimum. The enhanced gain relative to the case having no specific objective other than classification accuracy can be substantial.

## 7. Conclusions

Lapse risk is the most significant risk associated with life insurance. Lapses may cause losses, reduce expected profits, lead to stringent liquidity, result in mis-pricing, impair the risk management, or even pose solvency threats. Employing a good algorithm to model policyholder lapse behavior is therefore valuable.

In this study, we adopt innovative viewpoints on lapse management in addition to introducing machine learning algorithms to lapse prediction. Applying XGBoost and SVM to predicting whether a policyholder will lapse her/his policy is new to the literature. Secondly, we adopt not only a statistical metric in evaluating algorithms' prediction performance but also an economic metric based on customer lifetime value and retention gains.

The goal of classification accuracy has no direct link to the insurer's costs and profits. It thus might lead to a biased strategy (Powers, 2011). Following the churn literature, we define a specific validation metric based on the economic gains. This constitutes our third contribution: we are the first to set up a profit-based loss function so that we may directly optimize the economic gains. More specifically, we change the usual statistical idea of classification to a gain regression in which profits are to be maximized.

The two machine learning algorithms, XGBoost and SVM, perform a little bit better than classic CART and logistic regression in terms of statistical accuracy on a large dataset consisting of more than six hundred thousand life insurance policies with information on policy terms and policyholders' characteristics. XGBoost has another advantage over other algorithms: it is less dependent upon the choice of training samples.

The advantages of XGBoost and SVM are more apparent with respect to retention gains. The retention gains incorporate the costs of providing incentives to policyholders to reduce lapse propensities and the benefits of retaining policies. XGBoost and SVM generate much higher retention gains than logistic regression and CART do. For instance, XGBoost produces 1.2 to 2.6 million dollars more economic gains than CART.

In the last section, we demonstrate that the economic gains can be further enhanced when the optimization is done on a function linked to economic gains rather than on statistic accuracies. The results show that the retention gains with an aggressive incentive strategy resulted from XGB_R1 is 126% of those from applying XGBoost to pursue classification accuracies, in particular by reducing the false alarm rates. An insurer should therefore apply advanced machine learning algorithms like XGB to its economic objective so that lapse management can be really optimized.

## 9. Appendices

### 9.1. XGBoost Tuning – Binary Classification

The values of the parameters tested in the grid search for the tuning of XGBoost are as follows:

- *eta*: 0.05, 0.1, 0.15;
- *gamma*: 0, 5, 10;
- *max_depth*: 10, 15, 20, 25, 30;
- *min_child_weight*: 15, 20, 25;
- *subsample*: 1;
- *colsample_bytree*: 0.4, 0.5, 0.6.

The values of the grid search are chosen by a previous sensitivity study in which we apply the same methodology on a subsample of the whole database but with a coarser grid. Then we focus on a finer grid to obtain better results within a reasonable time period. In addition, the fact that we only test *subsample* with the value of 1 means that we do not adopt the stochastic gradient boosting of Friedman (2002).

### 9.2. SVM Tuning

The values of the parameters tested in the grid search for the tuning of SVM are as follows:

- *Cost*: 0.5, 1, 2, 5, 10;
- *gamma*: 0.25, 0.5, 0.75, 1, 1.25;

Similar to the previous section, the values of the grid search are chosen by a previous sensitivity study in which we apply the same methodology on a subsample of the whole database but with a coarser grid. Then we focus on a finer grid to obtain better results. This is necessary so that the computing can be done within a reasonable time period.

### 9.3. XGBoost Tuning – Profitability

We adopt the values of most parameters generated by a previous sensitivity study as:

- *eta* = 0.005;
- *gamma* = 1;
- *max_depth* = 15;
- *min_child* = 15;
- *subsample* = 0.7;
- *colsample* = 0.8.

Then, we determine the best *nrounds* through a 5-folds cross-validation with this parameter tested up to 1,000.